\documentclass[10pt,a4paper,reqno]{amsart}
\usepackage{amsthm,amsmath,amsfonts,amssymb,amsxtra,appendix,bookmark,dsfont,latexsym,bm,hyperref,delarray,color,euscript,amsgen,amsbsy,amsopn,amscd,latexsym,mathrsfs}
\usepackage[active]{srcltx}

\makeatletter

\usepackage{hyperref}
\makeatletter
\def\@tocline#1#2#3#4#5#6#7{\relax
  \ifnum #1>\c@tocdepth % then omit
  \else
    \par \addpenalty\@secpenalty\addvspace{#2}%
    \begingroup \hyphenpenalty\@M
    \@ifempty{#4}{%
      \@tempdima\csname r@tocindent\number#1\endcsname\relax
    }{%
      \@tempdima#4\relax
    }%
    \parindent\z@ \leftskip#3\relax \advance\leftskip\@tempdima\relax
    \rightskip\@pnumwidth plus4em \parfillskip-\@pnumwidth
    #5\leavevmode\hskip-\@tempdima
      \ifcase #1
       \or\or \hskip 1em \or \hskip 2em \else \hskip 3em \fi%
      #6\nobreak\relax
      \dotfill
      \hbox to\@pnumwidth{\@tocpagenum{#7}}
    \par
    \nobreak
    \endgroup
  \fi}
\makeatother

\newtheorem*{theorem*}{Theorem}

\theoremstyle{definition}

\theoremstyle{remark}

\newcommand\R{{\ensuremath {\mathbb R} }}
\newcommand\C{{\ensuremath {\mathbb C} }}

\newcommand\Z{{\ensuremath {\mathbb Z} }}

\renewcommand\phi{\varphi}
\newcommand{\bH}{\mathbb{H}}

\newcommand{\bW}{\mathbb{W}}

\newcommand{\gF}{\mathfrak{F}}

\newcommand{\cS}{\mathcal{S}}

\newcommand{\cF}{\mathcal{F}}
\newcommand{\cN}{\mathcal{N}}

\newcommand{\cZ}{\mathcal{Z}}

\newcommand{\bT}{\mathbb{T}^2}

\renewcommand{\epsilon}{\varepsilon}

\DeclareMathOperator{\tr}{{\rm Tr}}

\renewcommand{\ge}{\geqslant}

\renewcommand{\geq}{\geqslant}
\renewcommand{\leq}{\leqslant}
\renewcommand{\hat}{\widehat}

\newcommand{\bp}{\mathbf{p}}
\newcommand{\bq}{\mathbf{q}}
\newcommand{\bk}{\mathbf{k}}
\newcommand{\ak}{a_{\bk}}
\newcommand{\aka}{a_{\bk} ^{\dagger}}
\newcommand{\ap}{a_{\bp}}
\newcommand{\aq}{a_{\bq}}
\newcommand{\FNL}{\mathcal{E}^{\mathrm{int}}}

\begin{document}

\title{The interacting 2D Bose gas and nonlinear Gibbs measures}

\author[M. Lewin]{Mathieu LEWIN}
\address{CNRS \& CEREMADE, Universit\'e Paris-Dauphine, PSL Research University, Place de Lattre de Tassigny, 75016 Paris, France} 
\email{mathieu.lewin@math.cnrs.fr}

\author[P.T. Nam]{Phan Th\`anh Nam}
\address{Department of Mathematics, LMU Munich, Theresienstrasse 39, 80333 Munich, Germany} 
\email{nam@math.lmu.de}

\author[N. Rougerie]{Nicolas ROUGERIE}
\address{Universit\'e Grenoble-Alpes \& CNRS,  LPMMC (UMR 5493), B.P. 166, F-38042 Grenoble, France}
\email{nicolas.rougerie@grenoble.cnrs.fr}

\date{May, 2018}

\maketitle

%\tableofcontents

During the MFO workshop ``Gibbs measures for nonlinear dispersive equations'', we have announced a new theorem bearing on high-temperature 2D Bose gases. The purpose of this note is to state the result in a concise manner. Background, details, generalizations, discussion, references and proofs will appear elsewhere shortly. 

\medskip

\noindent\textbf{Hilbert space and state space}.  We consider the grand-canonical picture of the homogeneous 2D Bose gas. We assume periodic boundary conditions and thus particles live in the 2D unit flat torus~$\bT$. The particle number is not fixed: we work in the bosonic Fock space 
\begin{equation}\label{eq:Fock}
\gF = \C \oplus L^2 (\bT) \oplus \ldots \oplus L^2_{\rm sym} \left(\left(\bT\right) ^n\right) \oplus \ldots 
\end{equation}
where $L^2_{\rm sym} \left(\left(\bT\right) ^n\right)$ is the usual $n$-particle bosonic space of symmetric square-integrable wave-functions (identified with the $n$-fold symmetric tensor product of $L^2 (\bT)$ with itself).

We denote 
\begin{equation}
 \cS \left(\gF\right) := \left\{ \Gamma \mbox{ self-adjoint operator on } \gF, \: \Gamma \geq 0, \: \tr_{\gF} [\Gamma] = 1 \right\}
\end{equation}
the set of all (mixed) quantum states on the bosonic Fock space $\gF$. For any state $\Gamma \in \cS (\gF)$ of the form 
$$ \Gamma = \Gamma_0 \oplus \Gamma_1 \oplus \ldots \oplus \Gamma_n \oplus \ldots $$
we define its reduced $k$-body density matrix, a positive trace-class operator on $L_{\rm sym}^2 \left(\left(\bT\right) ^k\right)$, by the formula 
$$ \Gamma ^{(k)}:= \sum_{n\geq k} {n \choose k} \tr_{n+1\to k} \left[ \Gamma_n \right].$$
The partial trace $\tr_{n+1\to k}$ is taken on $n-k$ variables, no matter which by symmetry.

\medskip

\noindent\textbf{Hamiltonian}. We are interested in the equilibrium states of
\begin{equation}\label{eq:Hamil}
\bH_\lambda = \bH_0 + \lambda \bW, \quad  \bH_0 = \bigoplus_{n\geq 1} \sum_{j=1} ^n -\Delta_j  
\end{equation}
with $\lambda > 0$ a coupling constant and
$$ \bW = \bigoplus_{n\geq 2} \: \sum_{1\leq i < j \leq n} w (x_i - x_j), \quad \hat{w} \geq 0$$
where $w:\bT \mapsto \R$ is even and $\hat{w}$  is its Fourier transform (sequence of its Fourier coefficients). Equivalently, 
$$ \bH_\lambda = \sum_{\bk} |\bk| ^2 \aka \ak + \frac{\lambda}{2} \sum_{\bk,\bp,\bq} \hat{w} (\bk) a_{\bp + \bk} ^\dagger a_{\bq - \bk} ^\dagger \ap \aq $$
with annihilation $\ak$ and creation $\aka$ operators associated to the Fourier modes $e^{i \bk \cdot x}$, annihilating/creating a particle with momentum $\bk\in \left(2\pi \Z\right)^2$.

\medskip

\noindent\textbf{Quantum Gibbs state}. We investigate the minimizer, amongst states $\Gamma \in \cS (\gF)$, of the free-energy functional at temperature $T$ and chemical potential $\nu$, setting an energy reference $E_0$:
\begin{equation}\label{eq:free ener f}
\cF_{\lambda,T} [\Gamma] := \tr_{\gF} \left[ \left(\bH_\lambda - \nu \cN\right) \Gamma \right] + T \tr_{\gF} \left[ \Gamma \log \Gamma \right] + E_0.
\end{equation}
Here $\cN = \bigoplus_{n\geq 0} n = \sum_{\bk} \aka \ak$ is the particle number operator. The minimum free-energy is achieved by the Gibbs state 
\begin{equation}\label{eq:Gibbs}
\Gamma_{\lambda,T}:= \frac{1}{\cZ_{\lambda,T}} \exp\left( -\frac{1}{T} \left(\bH_\lambda - \nu \cN\right) \right) 
\end{equation}
where the partition function $\cZ_{\lambda,T}$ fixes the trace equal to $1$. The minimum free-energy is then 
$$ F_{\lambda,T} = - T \log \cZ_{\lambda,T} + E_0.$$

\medskip

\noindent\textbf{Nonlinear Gibbs measure}. Let $\kappa>0$ and $\mu_0$ be the gaussian measure with covariance $\left(-\Delta + \kappa\right) ^{-1}$. This is a probability measure supported on the negative Sobolev spaces $\bigcap_{s<0} H^s (\bT)$. Let $P_K$ be the orthogonal projector on the span of the Fourier modes with $|\bk| \leq K$. Consider then an interaction energy with local mass renormalization
\begin{equation}\label{eq:def int}
 \FNL_K [u] =  \frac{1}{2} \iint_{\bT \times \bT} \left( |P_K u(x)| ^2 - \left\langle |P_K u(x)| ^2 \right\rangle_{\mu_0} \right) w (x-y) \left( |P_K u(y)| ^2 - \left\langle |P_K u(y)| ^2 \right\rangle_{\mu_0} \right) dx dy. 
\end{equation}
Here $\left\langle \: . \: \right\rangle_{\mu_0}$ denotes expectation in the measure $\mu_0$. One can show that the sequence $\FNL_K [u]$ converges to a limit $\FNL[u]$ in $L^1 (d\mu_0)$ and that 
\begin{equation}\label{eq:NL measure}
d\mu (u) := \frac{1}{z} \exp\left(- \FNL [u] \right) d\mu_0 (u),  
\end{equation}
with $0<z<\infty$ a normalization constant, makes sense as a probability measure.

\medskip

\noindent\underline{\textbf{Result: the high-temperature/mean-field limit}}. Let $\kappa >0$ and denote 
$$ N_0 (T) := \sum_{\bk \in (2\pi\Z) ^2} \frac{1}{e^{\frac{|k|^2 + \kappa}{T}} - 1} $$
the expected particle number of the non-interacting quantum Gibbs state ($\lambda = 0$) at temperature $T$ and chemical potential $-\kappa$. This number is easily seen to be of order $T \log T$ for large $T$ and fixed~$\kappa$. Assume that 
$$ \hat{w} (\bk) \geq 0 \mbox{ for all } \bk \in (2\pi\Z) ^2 \mbox{ and } \sum_{\bk} \left(1 + |\bk| ^2\right) ^{1/2} \hat {w} (\bk) < \infty.$$
Then, we have the following 

\begin{theorem*}[\textbf{High-temperature/mean-field limit of the 2D Bose gas}]\mbox{}\\
Set, for some $\kappa >0$, 
\begin{equation}\label{eq:set chem}
 \nu =  \hat{w} (0) \lambda N_0 (T) - \kappa \quad \mbox{ and } \quad E_0 = \frac{1}{2} \lambda \hat{w} (0) N_0 (T) ^2.
\end{equation}
Then, in the limit $T\to \infty, \lambda T \to 1$ we have 
\begin{equation}\label{eq:CV free ener}
\frac{F_{\lambda,T} - F_{0,T}}{T} \to -\log z.
\end{equation}
Moreover, for every $k\ge 1$ and $p>1$
\begin{equation}\label{eq:CV DM Sp}
\tr  \left| \frac{k!}{T^k} \Gamma_{\lambda,T} ^{(k)} - \int |u^{\otimes k} \rangle \langle u^{\otimes k} | d\mu(u) \right| ^p \to 0. 
\end{equation}
Finally
\begin{equation}\label{eq:CV DM S1}
\tr \left| \frac{1}{T} \left(  \Gamma_{\lambda,T}^{(1)} -  \Gamma_{0,T} ^{(1)} \right) - \int |u \rangle \langle u | \left(d\mu(u) - d\mu_0(u)\right) \right| \to 0. 
\end{equation}
\end{theorem*}

\medskip

\noindent\textbf{Comments.} A detailed discussion is postponed to a future paper, that will also contain the proof of the theorem. The following remarks are thus intentionally kept to a bare minimum.

% \smallskip 
% 
% \noindent\textbf{1.} Physically this means that, modulo a precise tuning of the chemical potential, the quantum many-body problem for 2D bosons converges for large temperatures to a classical field theory. We expect but cannot prove yet that this is also true in 3D. 

\smallskip 

\noindent\textbf{1.} The 1D analogue of this theorem was proved first in~\cite{LewNamRou-14d}, see also~\cite{LewNamRou-17} and~\cite{FroKnoSchSoh-16,FroKnoSchSoh-17}. No renormalization is necessary to define the limit object in this case. The 2D and 3D cases are investigated in~\cite{FroKnoSchSoh-16} where the analogue of the above result is proved for some modified Gibbs state instead of the minimizer of the free-energy functional. 

\smallskip 

\noindent\textbf{2.} The construction of the nonlinear Gibbs measure $\mu$ requires renormalization because the natural interaction 
$$\frac{1}{2} \iint_{\bT \times \bT} |u(x)| ^2 w (x-y) |u(y)| ^2  dx dy$$ 
does not make sense on the support of the gaussian measure. The renormalized version~\eqref{eq:def int} is relatively simple to control because $\hat{w} \geq 0$. Positivity of the interaction is then preserved: $\FNL_K [u] \geq 0$ for all $u$. In more involved cases one can rely on tools from constructive quantum field theory, see~\cite{DerGer-13,GliJaf-87,Simon-74,VelWig-73} for reviews. 
%In 2D, replacing products by Wick products and using Nelson's method is usually sufficient.  

\smallskip 

\noindent\textbf{3.} Gibbs measures related to $\mu$ are known~\cite{Bourgain-96,Bourgain-97,OhTho-15} to be invariant under suitably renormalized nonlinear Schr\"odinger flows. They also appear as long-time asymptotes for stochastic nonlinear heat equations, see~\cite{MouWeb-15,RocZhuZhu-16,TsaWeb-16} and references therein for recent results. 

\smallskip 

\noindent\textbf{4.} The above theorem is part of the more general enterprise of gaining mathematical understanding on positive-temperature equilibria of the interacting Bose gas. The ground state and mean-field dynamics of this system are now well-understood, but rigorous works showing the effect of temperature seem rather rare~\cite{BetUel-10,DeuSeiYng-18,Seiringer-06,Seiringer-08,SeiUel-09,Yin-10}.

\smallskip 

\noindent\textbf{5.} In the physics literature, classical field theories~\cite{ZinnJustin-89} of the type we rigorously derive are used as effective descriptions at criticality, i.e. aroung the BEC phase transition, to obtain the leading order corrections due to interaction effects~\cite{ArnMoo-01,BayBlaiHolLalVau-99,BayBlaiHolLalVau-01,HolBay-03,KasProSvi-01}. Results of these papers are not easy to relate to our theorem, in particular because we work in 2D where there is no phase transition in the strict sense of the word. However~\eqref{eq:CV DM S1} is reminiscent of methods for calculating the critical density/critical temperature of the Bose gas in presence of interactions.    

\medskip

\noindent\textbf{Acknowledgements.} It is a pleasure to thank J\"urg Fr\"ohlich, Markus Holzmann, Antti Knowles, Benjamin Schlein, Vedran Sohinger and Jakob Yngvason for helpful discussions. Special thanks also to Giuseppe Genovese, Benjamin Schlein and Vedran Sohinger for organizing the workshop where the above result was first announced, and to the Mathematisches Forschungsinstitut Oberwolfach for hosting it. This project has received funding from the European Research Council (ERC) under the European Union's Horizon 2020 research and innovation programme (grant agreements MDFT No 725528 and CORFRONMAT No 758620).

%%%%%%%%%%%%%%%%%%%%%%%%%%%%%%%%%%%%%
% 
% \bibliographystyle{siam}
% \bibliography{/home/rougerie/Travail/Documentation/Bibtex/biblio-NR_Mai17}

\end{document}